\documentclass[doublecol,letterpaper]{epl2}
\usepackage{amsmath} 
\usepackage{amssymb} 
\usepackage{graphicx}
\usepackage{color}
\usepackage{url}
\usepackage{array}
\usepackage{txfonts}

\title{Rossby rogons in atmosphere and in the  solar photosphere}
\author{A. P. MISRA\inst{1}\thanks{E-mail: \email{apmisra@gmail.com}}
 \and P. K. SHUKLA\inst{2,3}\thanks{E-mail: \email{profshukla@yahoo.de}}
}
\institute{\inst{1} Department of Mathematics, Visva-Bharati University, Santiniketan-731 235, West Bengal, India\\
\inst{2} International Centre for Advanced Studies in Physical Sciences \& Institute for Theoretical Physics, Faculty of Physics and Astronomy, Ruhr University Bochum, D-447 80 Bochum, Germany\\
\inst{3} Department of Mechanical and Aerospace Engineering \& Center for Energy Research, University of California San Diego, La Jolla, CA 92093, U. S. A.}

\abstract{The   generation of Rossby rogue waves (Rossby rogons),   as well as the excitation of  bright and dark Rossby envelpe solitons are demonstrated on the basis of the modulational instability (MI) of a coherent Rossby wave packet. The evolution of an amplitude modulated Rossby wave packet is  governed by one-dimensional   (1D) nonlinear Schr\"odinger equation (NLSE).  The latter is   used to study the amplitude modulation of   Rossby wave packets  for fluids in   Earth's atmosphere and  in the solar photosphere.  It is found that an ampitude modulated  Rossby wave packet   becomes stable (unstable)  against quasi-stationary, long wavelength (in comparision with the Rossby wave length) perturbations, when the carrier Rossby wave number satisfies  $k^2 < 1/2$ or $\sqrt{2}+1<k^2<3$ ($k^2 >3$ or $1/2<k^2<\sqrt{2}+1$). It is also shown that   a Rossby rogon or a bright Rossby envelope soliton may be excited  in the shallow water approximation for the Rossby waves    in   solar photosphere. However, the excitation of small or large scale perturbations may be possible for magnetized plasmas in the ionosphereic $E-$layer.} 
\begin{document}
\maketitle
\section{Introduction}
The existence of rogue waves in oceanography \cite{ocean-rogue} and in nonlinear optics \cite{optics-rogue} has been known. Furthermore,  the appearance of such localized wave packets in  atmosphere \cite{atmosphere-rogue} as well as in plasmas \cite{plasma-rogue} has been a topic important research during the last few years. The physics of Rossby waves in shallow rotating fluids placed in a gravitational field is of interest owing not to its numerous theoretical as well as experimental studies \cite{rossby-th-exp1,rossby-th-exp2,rossby-th-exp3,rossby-th-exp4}, but also many important applications in astrophysics, space physics, physics of ocean and planetary atmosphere \cite{rossby-book-atmosphere}. 

The theory of magnetized Rossby waves in weakly ionized $E$ layer plasmas has been developed by Kaladze \textit{et al} \cite{rossby-E-layer}. They derived a generalized  Charney-Obukhov wave equation which contains both the scalar and vector nonlinearities. The latter has been shown to  balance  the wave dispersion  for the self-organization of solitary structures in magnetized plasmas. Recently, Kaladze and Wu \cite{rossby-photosphere} had shown that  Charney-Obukhov model can describe only the propagation of small-scale Rossy wave perturbations in the solar photosphere. On the other hand, the amplitude modulation of drift wave packets in a nonuniform magnetoplasma has been studied by Shukla and Misra \cite{drift-wave} owing to its important applications in laboratory and space environments. It was shown that such nonuniform magnetoplasmas with equilibrium electron density, electron temperature, and magnetic field inhomogeneities can support the existence of drift rogons, as well as bright and dark drift wave envelope solitons due to the amplitude modulation of a constant amplitude drift wave packet.
 
{In this letter, we study the  amplitude modulation of a coherent Rossby wave packet which may propagate  in   Earth's atmosphere and  in the solar photosphere. In contrast to two-dimensional Charney-Obukhov wave equation \cite{rossby-E-layer,rossby-photosphere} where the Jacobian (vector nonlinearity) plays important roles, we consider the one-dimensional propagation of Rossby waves along one of the horizontal directions (the $x$-direction)}. The model equation is considered in two different cases: one for fluids in the solar photosphere \cite{rossby-photosphere} and the other for weakly ionized $E$ layer plasmas \cite{rossby-E-layer}. We show that due to the modulational instability (MI) of Rossby waves, small-scale perturbations may develop into Rossby wave rogons or bright envelope Rossby solitons, which may be excited in the solar photosphere.  However, both the small and large scale perturbations of the Rossby waves may lead to the formation of rogons as well as bright and dark Rossby wave envelope solitons in magnetized ionospheric $E$ layer plasmas.

\section{Evolution equation and modulational instability}
We begin our discussion by considering the following one-dimensional Charney-Obukhov equation which describes the dynamics of solitary Rossby waves {(See for two-dimensioanl wave equation, e.g., Ref. \cite{rossby-E-layer,rossby-photosphere}):}  

\begin{eqnarray}
&& \frac{\partial}{\partial t}\left(h-\frac{\partial^2h}{\partial x^2}\right)+v_R \frac{\partial h}{\partial x}
 +v_R h \frac{\partial h}{\partial x}=0, 
\label{Charney-eq}
\end{eqnarray} 
where $h=\delta H/H_0<1$, $\delta H$ is the perturbation of the total depth of the fluid and $H_0$ is the uniform thickness in its stationary state. Also,  $v_R=-\beta r^2_R/c_g$ is the Rossby velocity normalized 
by the gravity wave speed $c_g=\sqrt{gH_0}$ ($g$ is the acceleration due to gravity), $\beta$ $(=\partial f/\partial y$ at $y=0)$ is the so-called Rossby parameter with $f$ denoting the Coriolis parameter. 
The space $x$ and time $t$ are normalized by the Rossby radius $r_R$ and $1/f_0\equiv r_R/c_g$, respectively. Equation \eqref{Charney-eq} may be considered as a model  for  the propagation of only 
small-scale perturbations in  solar photosphere \cite{rossby-photosphere} or for solitary waves in   ionospheric magnetized $E$ layer plasmas. 

We now derive the governing nonlinear equation for amplitude-modulated Rossby wave packets using the standard reductive perturbation technique \cite{reductive-1,reductive-2,reductive-3}. In the latter, one stretches the space and time variables as  $\xi=\epsilon (x-v_gt)$, $\tau=\epsilon^2 t$, where $\epsilon$ 
is a small parameter ($0 < \epsilon \ll 1)$ representing the weakness of perturbation and $v_g$  the Rossby wave group velocity to be obtained shortly.  In the  modulation of a plane Rossby wave as the carrier wave with the wave number $k$ and the frequency $\omega$,  the   variable  $h$ can be expanded as 
\cite{drift-wave}

\begin{equation} 
h=\sum^{\infty}_{n=1}\epsilon^n \sum^{\infty}_{l=-\infty}h_l^{(n)}(\xi,\tau)\exp[i(k x-\omega t)], 
\label{expansion}
\end{equation} 
where $h_{-l}^{(n)}=h_{-l}^{(n)*}$ is the reality condition and the asterisk denotes the complex conjugate.

Substituting the expansion \eqref{expansion} and the aforementioned  stretched coordinates into Eq. \eqref{Charney-eq}, and equating different  powers of $\epsilon$, we obtain for $n=l=1$ (coefficient of $\epsilon$) the following linear dispersion relation for the Rossby waves
\begin{equation}
\omega =\frac{v_R k}{1+k^2}.
\label{DR}
\end{equation}

For $n=2,l=1$, i.e., for the second order first harmonic modes, we obtain an equation in which the coefficient of $h_1^{(2)}$ vanishes due to the usage of the Rossby wave dispersion relation.  Then, the coefficient of $\partial h_1^{(1)}/\partial\xi$, while equating to zero,  gives the  group velocity 
\begin{equation} 
v_g=\frac{v_R(1-k^2)}{(1+k^2)^2}. \label{vg}
\end{equation} 
Proceeding in this way we obtain, from the coefficients of $\epsilon^2$ and $\epsilon^3$,  the second and zeroth order  harmonic modes for $n=l=2$ and $n=2,l=0$ as  
\begin{equation}
h_2^{(2)}=\frac{v_R k}{\omega(1+4k^2)-2v_R k}[h_1^{(1)}]^2, \hskip5pth_0^{(2)}=\left(\frac{v_R}{v_g-v_R}\right)|h_1^{(1)}|^2. \label{2ndorder}
\end{equation}

Finally, for $n=3, l=1$,  we obtain an equation in which the coefficients of $h^{(3)}_1$ 
and $\partial h_1^{(2)}/\partial\xi$ vanish by the Rossby wave dispersion relation \eqref{DR} and the Rossby wave group velocity dispersion \eqref{vg}, respectively.  In the reduced equation, we substitute the expressions for $h_0^{(2)}$ and $h_2^{(2)}$ from Eq. \eqref{2ndorder}   to  obtain the following nonlinear Schr\"odinger equation (NLSE)
\begin{equation}
i\frac{\partial h}{\partial \tau}+P\frac{\partial^2 h}{\partial \xi^2}+Q|h|^2h=0, \label{NLSE}
\end{equation}
where we redefine $h\equiv h^{(1)}_1$, which may be of   the order of unity or less, so that in the expansion $h=\epsilon h^{(1)}_1+\cdots$ [Eq. \eqref{expansion}],  $h\ll1$.

In Eq. \eqref{NLSE}, the group velocity  dispersion and the nonlinear coefficients are, respectively,
\begin{equation}
P \equiv\frac{1}{2}\frac{\partial^2 \omega}{\partial k^2}= \frac{v_R k(k^2-3)}{(1+k^2)^3},
\label{P-dispersion}
\end{equation}
and
\begin{equation}
Q=\frac{v_R(\sqrt{2}+1-k^2)(\sqrt{2}-1+k^2)}{k(3+k^2)(1-2k^2)}, 
\label{Q-nonlinear}
\end{equation}

{It has been found that in the atmosphere, solar photosphere as well as in the ocean, the large scale flows can be caused by the nonlinear interaction of different types of small scale linear wave modes, e.g., Rossby waves. The modulation of   slowly varying wave amplitudes   occur due to  self-interaction of these carrier wave modes. These phenomena, however,  are relevant to the MI  which constitutes one of the most fundamental effects associated with the   propagation of wave packets in nonlinear media.  Such MI   signifies the exponential growth or decay of a small perturbation of the waves. The gain eventually leads to the amplification of sidebands, which break up the otherwise uniform wave, and lead to energy localization via the formation of localized structures. Thus, the system's evolution is then governed through the  MI, which  acts as a precursor for the formation of rogons or bright  envelope solitons. However, in absence of the MI, the stable  wave packet can propagate in the form of a dark envelope soliton. 

The amplitude modulation   of Rossby waves in the Earth's atmosphere due to the action of earthquake's electromagnetic radiation has been considered by Tsintsadze \textit{et al} \cite{MI_Rossby}. The generation of different stable and unstable branches of oscillations (depending on the intensity of the  pumping waves)   by seismic activity has been reported there. It has been predicted that such  oscillation branches along with energetically reinforcing Rossby solitary vortical anticyclone structures may serve as precursors to earthquakes.  However, in our nonlinear theory to be presented shortly we have considered the amplitude modulation of Rossby waves that may propagate in weakly ionized $E$ layer magnetoplasmas as well as in fluid media in the solar photosphere. We show that the amplitude modulation of a coherent Rossby wave in these environments leads to the localization of bright envelope solitons or the emergence of Rossby rogons. The latter may emerge in the planetary atmosphere as well as in the solar photosphere, and can play the roles of precursors of certain extraordinary natural phenomena (earthquakes, magnetic storms, volcano eruptions or man-made activities  etc.)
 }   

We now consider the amplitude modulation of a plane Rossby wave solution of Eq.  \eqref{NLSE} of the form $h=h_0e^{-i\Omega_0\tau}$, where $\Omega_0=-Qh_0^2$ and $h_0$ is a constant.  The Rossby wave amplitude is then modulated against a plane wave perturbation with    frequency $\Omega$ and   wave number $K$ as $h=\left(h_0+h_1e^{iK\xi-i\Omega\tau}+h_2 e^{-iK\xi+i\Omega\tau}\right)e^{-i\Omega_0\tau}$, where $h_{1,2}$ are real constants. Since the  perturbations are assumed to be small and nonzero, we obtain from Eq. \eqref{NLSE} the following dispersion relation for the modulated Rossby wave packets \cite{drift-wave}
\begin{equation}
\Omega^2=(PK^2)^2\left(1-\frac{K_c^2}{K^2}\right),\label{DR-modulation}
\end{equation}
where $K_c= \sqrt{2|Q/P|}|h_0|$ is the critical value of   $K$,  such that  the MI sets in for $K<K_c$, and the wave will be modulated for $PQ>0$.   On the other hand, for $K>K_c$ the Rossby wave is said to be stable $(PQ<0)$ against the modulation. The instability growth  rate is  given by
\begin{equation}
\Gamma= |P| K^2\sqrt{\frac{K^2_c}{K^2}-1}. \label{instability-rate}
\end{equation}
Clearly, the maximum value of $\Gamma$ is  achieved at $K=K_c/\sqrt{2}$, and is given by $\Gamma_{\text{max}}=|Q||\Phi_0|^2$. 

Inspecting on the expressions for $P$ and $Q$ given by Eqs. \eqref{P-dispersion}, \eqref{Q-nonlinear},    we find that   $PQ\gtrless0$ according as $(k^2-3)(\sqrt{2}+1-k^2)(1-2k^2)\gtrless0$. So, $PQ>0$  for either $k^2>3$ or  $1/2<k^2<\sqrt{2}+1$, and $PQ<0$ according to which  $k^2<1/2$ or $\sqrt{2}+1<k^2<3$.   We find that for the MI to occur and for a fixed $v_R$,     as $k$ increases   $(>\sqrt{3})$, the value of $\Gamma$ also increases with a  cut-off at higher   wave number of modulations. A similar trend occurs when $k$ also lies in $1/2<k^2<\sqrt{2}+1$ for which $PQ>0$.  This indicates that for long wavelength perturbations with $K<1$, the MI growth rate can  be controlled for Rossby waves with wave numbers $k$ close to either $1/\sqrt{2}$ or $\sqrt{3}$.  
Thus, we   conclude that the Rossby wave packets may be stable or unstable   against the  modulation at small $(k>1)$, as well as for large scale $(k<1)$ perturbations. As is evident from Ref. \cite{rossby-photosphere} that the model \eqref{Charney-eq} describes only the propagation of  small-scale low-frequency perturbations  in the  solar photosphere. However, Eq. \eqref{Charney-eq} may be used to describe  both the small as well as large scale phenomena for weakly ionized $E$-layer plasmas \cite{rossby-E-layer}. In our numerical investigation below, we will consider those cases separately, and show that in both the cases, the Rossby wave packet may propagate as a rogon, as well as bright or dark envelope solitons.        
\section{Solutions of the NLSE}
Exact solutions of the NLSE \eqref{NLSE} can be presented (see for details, e.g., Refs. \cite{soliton-solutions}). Assuming $h=\sqrt{\Psi}\exp(i\theta)$, where $\Psi$ and $\theta$ are real functions, the    bright envelope  soliton solution of Eq. \eqref{NLSE} is obtained for $PQ>0$ as \cite{drift-wave}
\begin{equation}
\Psi=\Psi_0 \hskip1pt\text{sech}^2\left(\frac{\xi-U\tau}{W}\right),\hskip2pt \theta=\frac{1}{2P}\left[U\xi+\left(\Omega_0-\frac{U^2}{2}\right)\tau\right].\label{bright-envelope}
\end{equation} 
This represents a localized pulse traveling at a speed $U$ with  oscillation frequency  $\Omega_0$ at rest. The pulse width $W$ is given by $\Psi_0$ as $W=\sqrt{2P/Q\Psi_0}$, where $\Psi_0$ is a constant.
\begin{figure}
\begin{center}
\includegraphics[height=3.4in,width=3.4in]{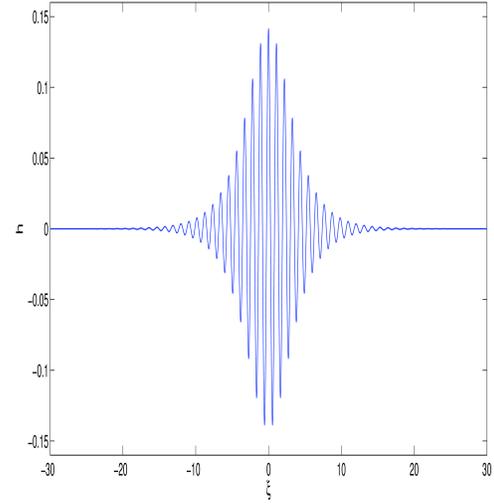}  
\caption{The evolution of the bright envelope soliton as given by Eq. \eqref{bright-envelope}  at $\tau=0$ for $k=5$,  $\Psi_0=0.02$, $\Omega_0=0.1$ and $U=0.2$.} 
\end{center}
\end{figure}

Furthermore,  Eq. \eqref{NLSE} can have a rogue wave solution that is located in a nonzero background and localized both in space and time,  given  by (for $PQ>0$) \cite{atmosphere-rogue,drift-wave}
\begin{equation}
h = h_0\left[\frac{4\left(1+2 i Q\tau\right)}{1+4Q^2\tau^2 + 2Q \xi^2/P}-1 \right]\exp\left(i Q\tau\right). \label{rogue-solution}
\end{equation}

On the other hand, for $P Q < 0$, the modulationally stable  wave packet will propagate in the form of a dark envelope soliton   characterized by a depression of the Rossby wave perturbation around $\xi =0$. This is given by \cite{drift-wave} 
\begin{eqnarray}
&&\Psi=\Psi_1\hskip1pt \text{tanh}^2\left(\frac{\xi-U\tau}{W_1}\right), \notag \\ &&\theta=\frac{1}{2P}\left[U\xi-\left(\frac{U^2}{2}-2PQ\Psi_1\right)\tau\right].\label{dark-envelope}
\end{eqnarray} 
 The solution \eqref{dark-envelope} represents a localized region of hole (void) traveling at a speed $U$. The pulse width $W_1$ depends on the constant amplitude $\Psi_1$ as $W_1=\sqrt{2|P/Q|/\Psi_1}$.
\section{Rossby waves in the solar photosphere}
Following  Ref. \cite{rossby-photosphere},  we consider  the low-frequency oscillations characterized by a small Rossby number, i.e.,   $\epsilon_{\omega}\sim (1/f)(\partial/\partial t)\sim (\omega/f)\ll1$. We also assume that    the length scale $L$ of motion is much smaller than the scale of variation of $f$, i.e., 
 $\epsilon_{\beta}\sim(Lf^{\prime}/f) \ll1$,  and that  the amplitude of oscillations,
 $\epsilon_h\sim h\ll1$. We note that the ratio of the parameters $\epsilon_{\omega}$ and $\epsilon_h$ gives a typical length scale,  $\epsilon_{\omega}/\epsilon_h\sim r_R^2/L^2$,
 which shows that if $h>\epsilon_{\omega}$, then $L$ must be larger than $r_R$.  In the small-scale dynamics $(k>1)$ assuming $k\sim \delta/r_R$, where $\delta>1$, we obtain (considering magnitude only) the following estimates for $\omega$ and $v_g$ (in dimensional forms) as 

\begin{equation}
\omega\sim\frac{\beta}{k}\sim\frac{\beta r_R}{\delta}; \hskip5pt v_g\sim\frac{\beta}{k^2}\sim\frac{\beta r^2_R}{\delta^2}.  \label{estimate-1}
\end{equation}

For the photospheric parameters \cite{rossby-photosphere} with  $H_0\approx500$ km, the solar surface gravity $g\approx274$ m$/s^2$, the Coriolis parameter $f\approx5.8$ $\mu$ Hz, for which  $r_R\approx2\times10^9$ m $>R_{\odot}$, the solar radius and taking $\delta=5$, we find from Eqs. \eqref{estimate-1}  that $\omega\sim3.2$ $\mu$Hz, $v_g\sim1.28\times10^3$ m/s. These estimates well agree with those obtained numerically and will be   appropriate to explain the observed data \cite{rossby-photosphere}. Furthermore, we also find that in the small scale approximation, $PQ$ is always positive, giving rise the possibility of the existence of  Rossby wave packets in the form of  bright envelope solitons  or rogons. The profile of the bright envelope solitons at $k=5$ is shown in Fig. 1. Figure 2 shows that the Rossby rogue wave can be generated at small scales. For smaller values of the Coriolis parameter, it is seen that the wave packet gets  localized in relatively a small space and time implying that a significant amount of energy is concentrated   in  a relatively small area in  space.  Thus, a random perturbation of the Rossby wave amplitude will grow on account of the modulational instability.

 \begin{figure*}
\begin{center}
\includegraphics[height=3.6in,width=7in]{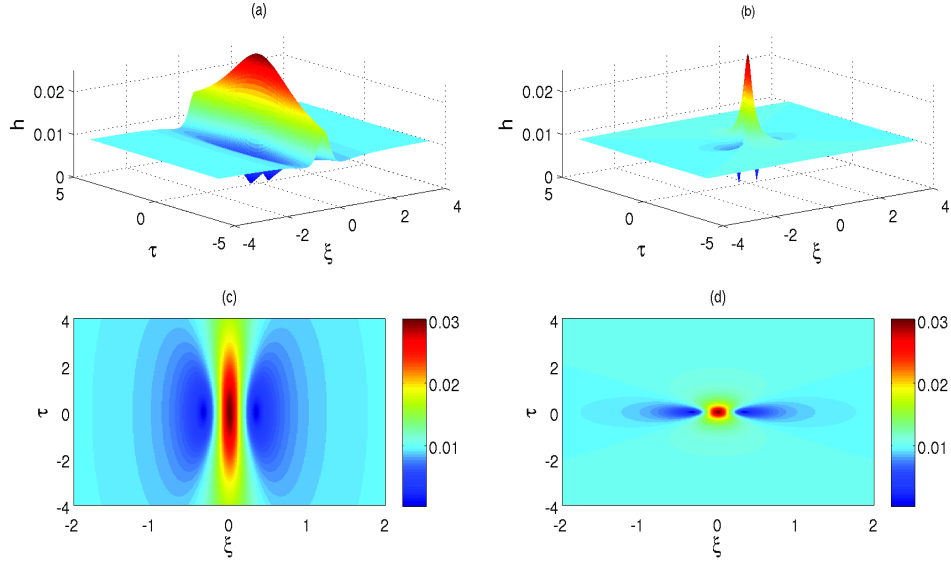}  
\caption{The evolution of the rogon as given by Eq. \eqref{rogue-solution} at $k=5$  with $h_0=0.01$. The plots (a) and (b)   are, respectively, for  $f=5.8$ $\mu$ Hz and $f=1.8$ $\mu$ Hz. The plots (c) and (d) are  contour plots corresponding to (a) and (b) respectively.} 
\end{center}
\end{figure*} 
\begin{figure*}
\begin{center}
\includegraphics[height=3.2in,width=7in]{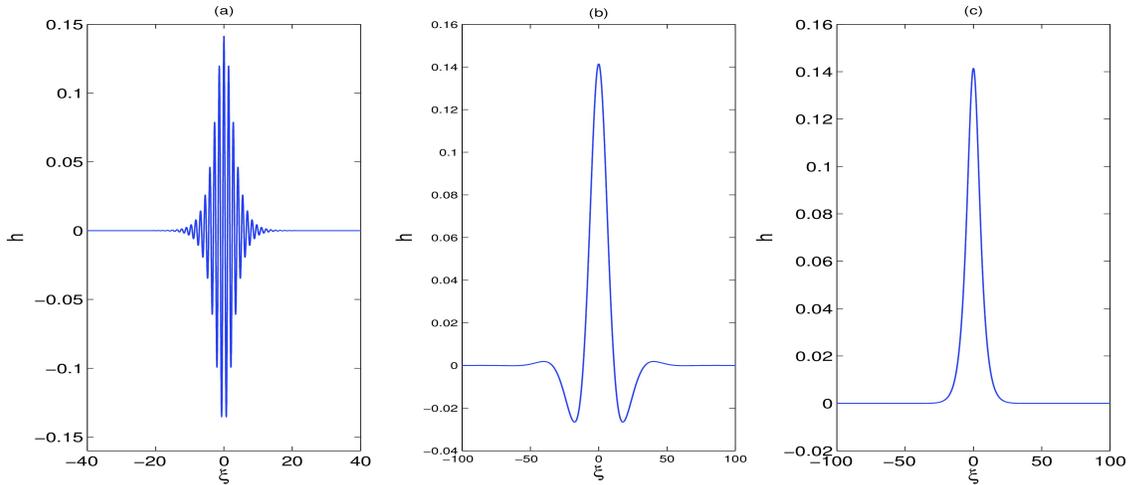}  
\caption{The evolution of  bright envelope soliton [Eq. \eqref{bright-envelope}] for Rossby wave packets in weakly ionized ionospheric $E$ layer plasmas at three different scales: (a) Small scale $(k>1)$ with  $k=5$, (b) Intermediate scale $(k\sim1)$  with $k=1.2$   and (c) Large scale $(k<1)$. Here    $\tau=0$,   $\Psi_0=0.02$, $\Omega_0=0.1$ and $U=0.2$. The other parameter values are as in the text.} 
\end{center}
\end{figure*}
\section{Rossby waves in the ionospheric $E$ layer} 
The model Eq. \eqref{Charney-eq} can be considered to describe the Rossby solitary wave dynamics for ionosphereic magnetized $E$ layer plasmas \cite{rossby-E-layer}.   In this case, the corresponding parameter values will be different (see for details Ref. \cite{rossby-E-layer}) from those considered in the case of solar photosphere. For illustration, we consider $H_0=200$ km, $\beta\sim1.1\times10^{-11}$ /ms, $f\sim5.2\times10^{-5}$ /s at latitude $\lambda=\pi/6$ and $r_R\sim2.6\times10^7$ m. { Here the Coriolis parameter $f$ is obtained using the $\beta$ plane approximation  in the vicinity of the latitude $\lambda=\lambda_0$, as $f\approx2\Omega_0\sin \lambda$, where $\Omega_0=50$ $\mu$Hz is the constant rotation frequency \cite{rossby-E-layer}. The Rossby radius is then obtained from the relation $r_R=\sqrt{gH_0}/f$, where $g$ is the acceleration due to gravity.}  We see that the generation of bright envelope (Fig. 3) and rogue wave solutions (Fig. 4) may be possible at three different scales: (i) small scale $(k>1)$ with $k=5$ for which $\omega\sim1.4$ $\mu$Hz, $v_g\sim1.03\times10^3$ m/s, (ii) intermediate scale $(k\sim1)$ with $k=1.2$ for which $\omega\sim1.4$ $\mu$Hz, $v_g\sim5.68\times10^2$ m/s  and (iii) large scale $(k<1)$ with $k=0.8$ for which $\omega\sim1.4$ $\mu$Hz, $v_g\sim1.03\times10^3$ m/s. From Fig. 3, it is also found that the Rossby wave packets propagate as single solitary pulses with reduced amplitudes at intermediate and large-scale motions. However, the dark envelope solitons (Fig. 5) may exist only at intermediate $(k=1.61)$ and large scale $(k=0.02)$  motions. 
 \begin{figure*}
\begin{center}
\includegraphics[height=3.8in,width=7in]{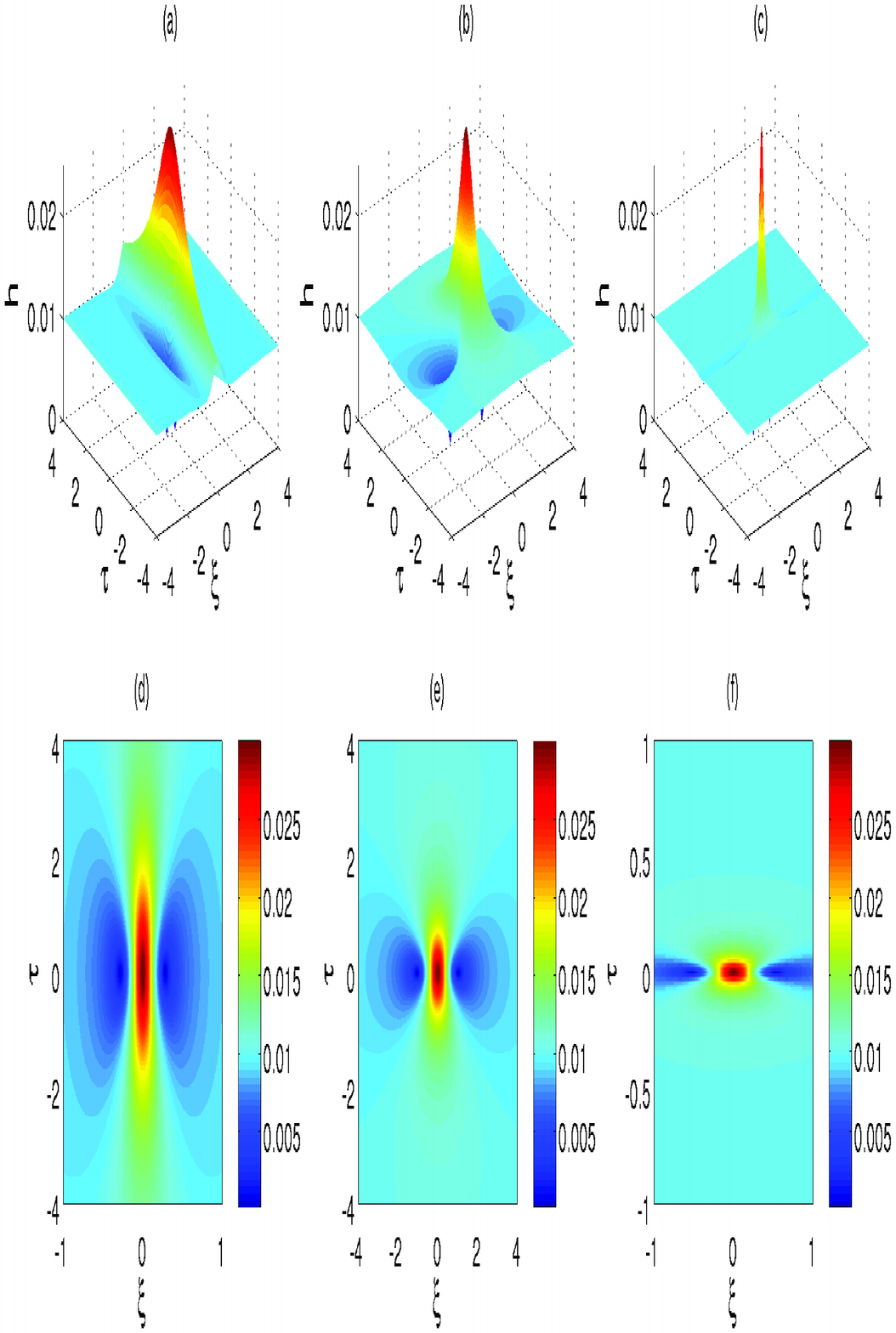} 
\caption{The evolution of rogons [Eq. \eqref{rogue-solution}] for Rossby wave packets in weakly ionized ionospheric $E$ layer plasmas at three different scales: (a) Small scale $(k>1)$ with  $k=5$, (b) Intermediate scale $(k\sim1)$  with $k=1.2$   and (c) Large scale $(k<1)$  with $k=0.8$. The  corresponding contour plots are shown in (d), (e) and (f) respectively. Here  $h_0=0.01$ and other parameter values are as in the text.  } 
\end{center}
\end{figure*} 
\begin{figure}
\begin{center}
\includegraphics[height=3.4in,width=3.4in]{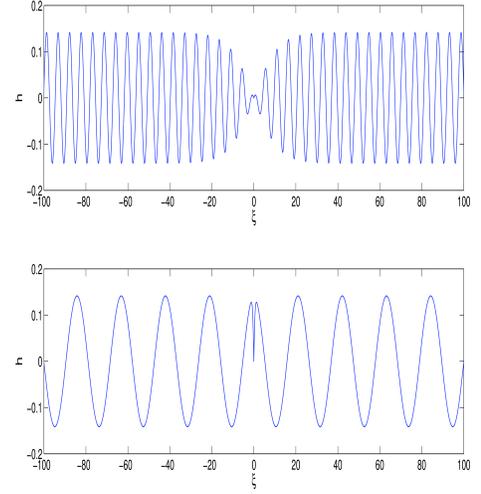} 
\caption{The evolution of  dark envelope solitons [Eq. \eqref{dark-envelope}] for Rossby wave packets in weakly ionized ionospheric $E$ layer plasmas  at two different values of $k$: $k=1.6$ (upper panel)  and $k=0.02$ (lower panel) corresponding to intermediate $(k\sim1)$ and large scale $(k<1)$ motions with  $\tau=0$, $\Psi_1=0.02$ and $U=0.2$. The other parameters are as in the text.} 
\end{center}
\end{figure}

\section{Conclusion} We have considered the amplitude modulation of   Rossby wave packets that may propagate  for fluids in the solar photosphere \cite{rossby-photosphere} as well as for weakly ionized magnetized plasmas in the ionospheric $E$ layer \cite{rossby-E-layer}.  We show that the generation of rogue waves (rogons) or bright/dark envelope solitons may be possible in these environments.  Starting from the one-dimensional Charney-Obukhov equation for the dynamics of Rossby waves, and using the standard reductive perturbation technique, we derive a nonlinear Schr\"odinger  equation which describes the dynamics of small amplitude Rossby wave packets. It is found that the wave packets may propagate as rogons or bright envelopes when   either $k^2>3$ or  $1/2<k^2<\sqrt{2}+1$.   On the other hand, the generation of dark  envelope soliton is also possible for either  $k^2<1/2$ or $\sqrt{2}+1<k^2<3$. Since the Charney-Obukhov model is valid for only small-scale low-frequency perturbations in the solar photosphere, the existence of dark envelopes may not be possible there, whereas both the bright and dark envelope solitons may be excited in the ionospheric $E$ layer plasmas. 

{It is to be mentioned that, in one-dimensional propagation, the vector or Poisson-bracket  nonlinearity can be neglected.  It can also be neglected in the large-scale dynamics of Rossby waves  in which the size $a$ of the soliton exceeds the Rossby radius $r_R$, i.e., $a\gg r_R$  \cite{rossby-E-layer}. Since $\xi$ is normalized by $r_R$, it is clear from Fig. 3 that the soliton width (size) exceeds $r_R$ and the inequality can be satisfied. Furthermore, in two-dimensional motion, since the vector nonlinearity involves the symmetric derivatives of 
$h$ and $\nabla^2h$ with respect to $x$ and $y$, the contributions  from the nonlinear interactions of the higher  harmonic modes (for $n=2,l=2$; $n=2, l=1$; $n=2, l=0$ and $n=3, l=1$) get cancelled.}   In conclusion, the present results should be helpful in identifying modulated  Rossby wave packets  { that may be generated} in the solar photosphere,  ionospheric $E$ layer, as well as laboratory fluids or plasmas.
\acknowledgements
{This research was partially supported by the Deutsche Forschungsgemeinschaft (DFG),
Bonn, through the project SH21/3-2 of the Research Unit 1048, and by the International
Space Science Institute in Bern (Switzerland) through the project ``Large-Scale Vortices and Zonal Winds in Planetarry Atmospheres/Ionospheres: Theory Versus Observations". APM acknowledges partial support from the SAP-DRS (Phase-II), UGC, New Delhi, through sanction letter No. F.510/4/DRS/2009 (SAP-I) dated 13 Oct., 2009, as well as from Visva-Bharati.}

\bibliographystyle{elsarticle-num}

\end{document}